\documentclass{appolb}
\usepackage{graphicx}
\usepackage{amsmath}
\usepackage{amsthm}
\usepackage{amsfonts}
\usepackage{amssymb}
\begin{document}
% \eqsec  % uncomment this line to get equations numbered by (sec.num)
\title{Hadronic Resonances in Lattice QCD%
\thanks{Presented at {\it Excited QCD}, 3-9 February 2013, Bjela\v snica Mountain, Sarajevo}%
% you can use '\\' to break lines
}
\author{S. Prelovsek$^{(a,b)\footnote{e-mail:sasa.prelovsek@ijs.si}}$, C. B. Lang$^{(c)}$, L. Leskovec$^{(b)}$, D. Mohler$^{(d)}$ and R. M. Woloshyn$^{(e)}$
\address{(a) Physics Department, University of Ljubljana, 1000 Ljubljana, Slovenia}
\address{(b) Jozef Stefan Institute, 1000 Ljubljana, Slovenia}
\address{(c) Institut f\"ur Physik,  Universit\"at Graz, A-8010 Graz, Austria}
\address{(d) Fermilab, Batavia, 60510-5011, Illinois, USA}
\address{(e) TRIUMF, 4004 Wesbrook Mall, Vancouver, BC V6T 2A3, Canada}
}
\maketitle
\begin{abstract}
I discuss how masses and widths of hadron resonances are extracted from lattice QCD. 
Recent lattice results on the light, strange and charm meson resonances are reviewed. Their properties are revealed by simulating the corresponding scattering channels $\pi\pi$, $K\pi $ and $D\pi$ on the lattice and extracting the scattering phase shifts. In particular we address the resonances $\rho$, $D_0^*(2400)$, $D_1(2430)$, $K^*$, $\kappa$ and $K_0^*(1430)$.
\end{abstract}
%\PACS{PACS numbers come here}
  
\section{Introduction}
Most  hadrons are resonances, i.e, they decay extremely fast via the strong interaction. Yet most of these resonances were studied in lattice QCD assuming the so-called narrow width approximation, that is ignoring  their strong decay. Up to now only the $\rho$ meson has been simulated properly as a resonance by several groups  and its width was extracted (see \cite{Lang:2011mn,Pelissier:2012pi,Dudek:2012xn} and references therein). Recently the first  simulation of  charmed resonances in $D\pi$ scattering was performed \cite{Mohler:2012na}, while the strange scalar and vector resonances were addressed by simulating $K\pi$ scattering \cite{Lang:2012sv}.  This paper briefly reviews the main results and methods that were employed.  Recent lattice results of resonances are reviewed in \cite{Mohler:2012nh}.  

\section{Meson-meson scattering in a resonant channel on the lattice}

Meson resonances are formed in the strong scattering  $M_1M_2\to R\to M_1M_2$ in partial wave $l$.  They exhibit a  Breit-Wigner-like resonance behavior of  $\delta$, amplitude $T$ and  $\sigma(s)\propto \sin^2\delta(s)$
\begin{align}
T=\frac{-\sqrt{s}\,\Gamma(s)}{s-m_R^2+i\sqrt{s}\,\Gamma(s)}&=e^{i\delta(s)}\sin \delta(s)~,\quad 
\Gamma(s)=\frac{(p^*)^{2 l+1}}{s} ~g^2~ \quad \label{BW} \\
 \frac{(p^*)^{2 l+1} ~ \cot\delta}{\sqrt{s}}&=\frac{1}{g^2}(s-m_R^2)\label{BW_xy} 
\end{align}
where $\Gamma(s)$ is parametrized in terms of the phase space and the $R\to M_1M_2$ coupling $g$.    The combination $ (p^*)^{2 l+1} \cot\delta/\sqrt{s}$ is linear in $s$ for a single Breit-Wigner resonance, which allows the extraction of $m_R$ and $g$ (and therefore the width) using a linear fit (\ref{BW_xy}) once the phase shifts $\delta(s)$ are determined from the lattice.  So the  goal is  to simulate the scattering on the lattice and determine the scattering phase shift $\delta(s)$.

For this purpose one computes the correlator $C_{ij}(t)=\langle 0|{\cal O}_i(t){\cal O}_j(0)|0\rangle$. 
We use the interpolators ${\cal O}=M_1(\vec p_1)M_2(\vec p_2)=\bar q_1\Gamma_1 q_1^\prime ~\bar q_2\Gamma_2 q_2^\prime$ that create two-meson states with definite momenta, and  ${\cal O}=\bar q \Gamma q^\prime$ that couple well to the resonances. Both are constructed to have the quantum numbers\footnote{On a finite  discrete lattice the interpolators ${\cal O}$ have to transform according to irreducible representations of the symmetry group related to the  center-of-momentum frame. }  of  the desired channel and total momentum $\vec P=\vec p_1+\vec p_2$.   
 The interpolators ${\cal O}$ couple in general to all physical eigenstates $n$ and each of them evolves as $e^{-E_nt}$  in euclidean time, so $C_{ij}(t)=\sum_n A_{ij}^{(n)}~ e^{-E_n t}$.
 
We calculate the correlators $C_{ij}(t)$  using a powerful  distillation method \cite{Peardon:2009gh}, which enables the calculation of all the necessary Wick contractions. Our study is based on  280 gauge configurations  with  $a\simeq 0.124~$fm  and dynamical Wilson-Clover $u/d$ quarks corresponding to $m_\pi\simeq 266$ MeV. 
A rather small volume $16^3\times 32$  makes the distillation method \cite{Peardon:2009gh} feasible. The valence charm quark is treated using the Fermilab method described in \cite{Mohler:2012na}. 
 The  resulting $C_{ij}(t)$  allows the extraction of the few lowest eigen energies  $E_n$ via the generalized eigenvalue method. 

\vspace{-0.5cm}

\section{Physics information based on the energy spectrum $E_n(L)$}

\begin{figure}[tbh]
\begin{center}
\includegraphics[width=55mm,clip]{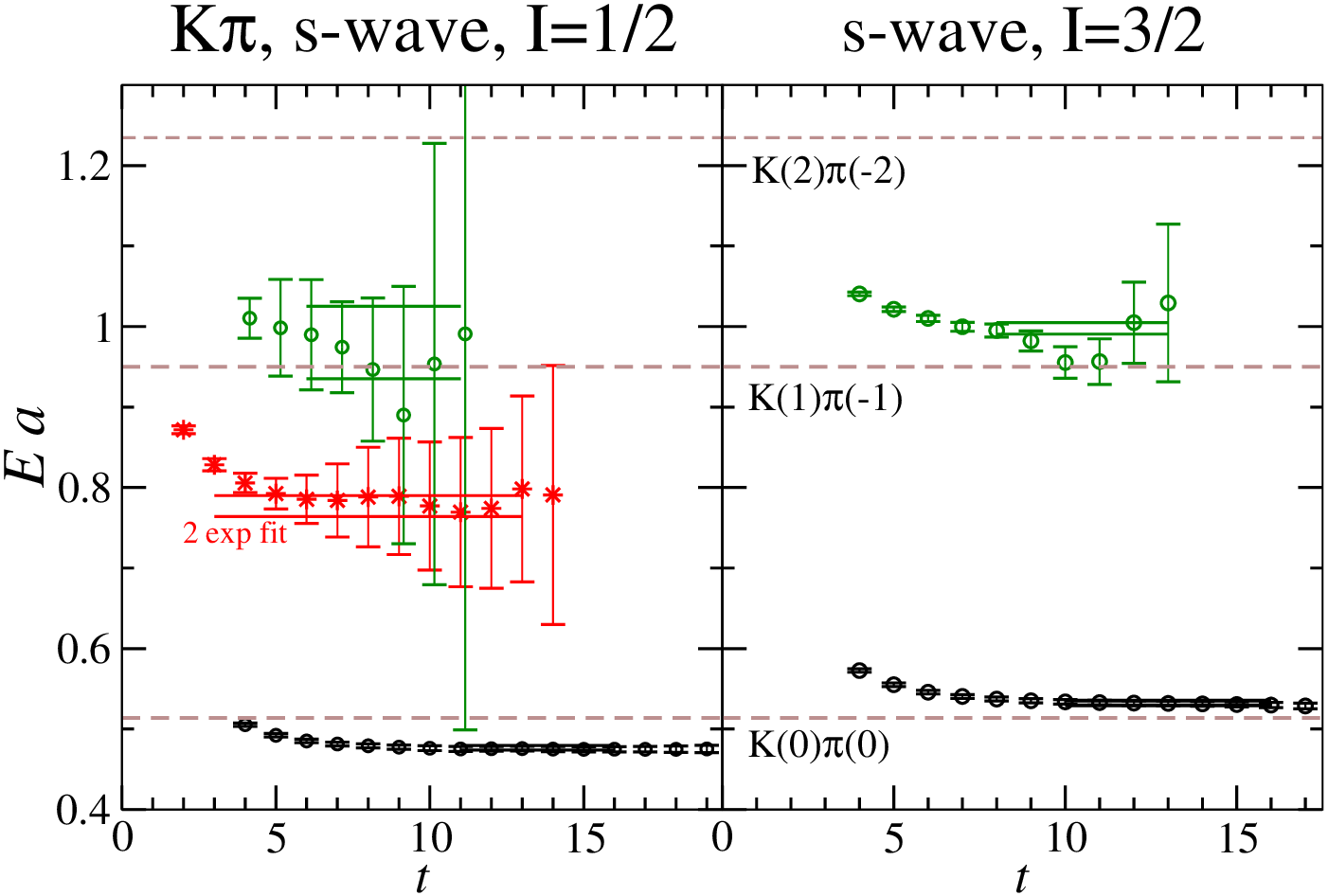}
\includegraphics[width=52mm,clip]{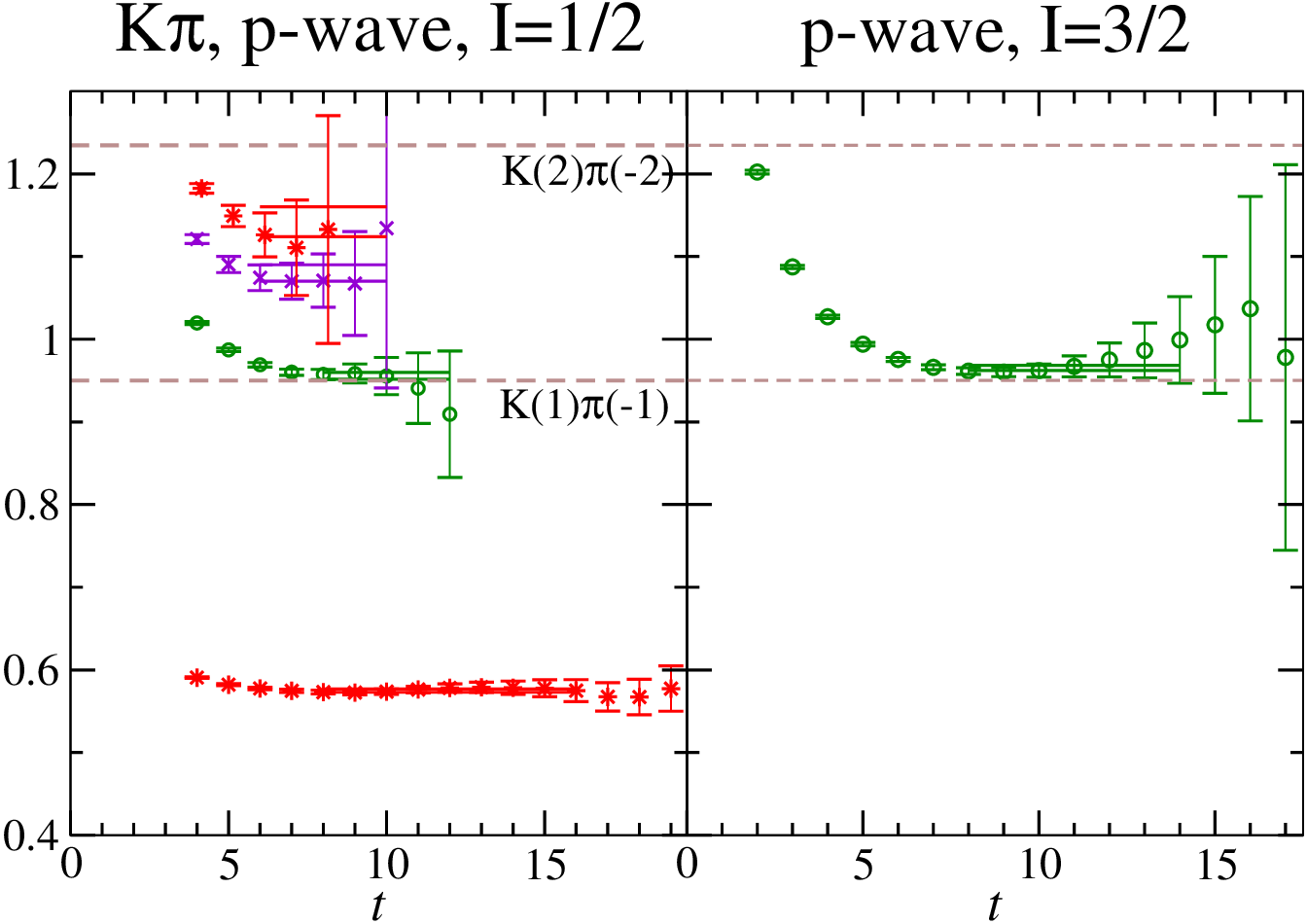}
\includegraphics[width=45mm,clip]{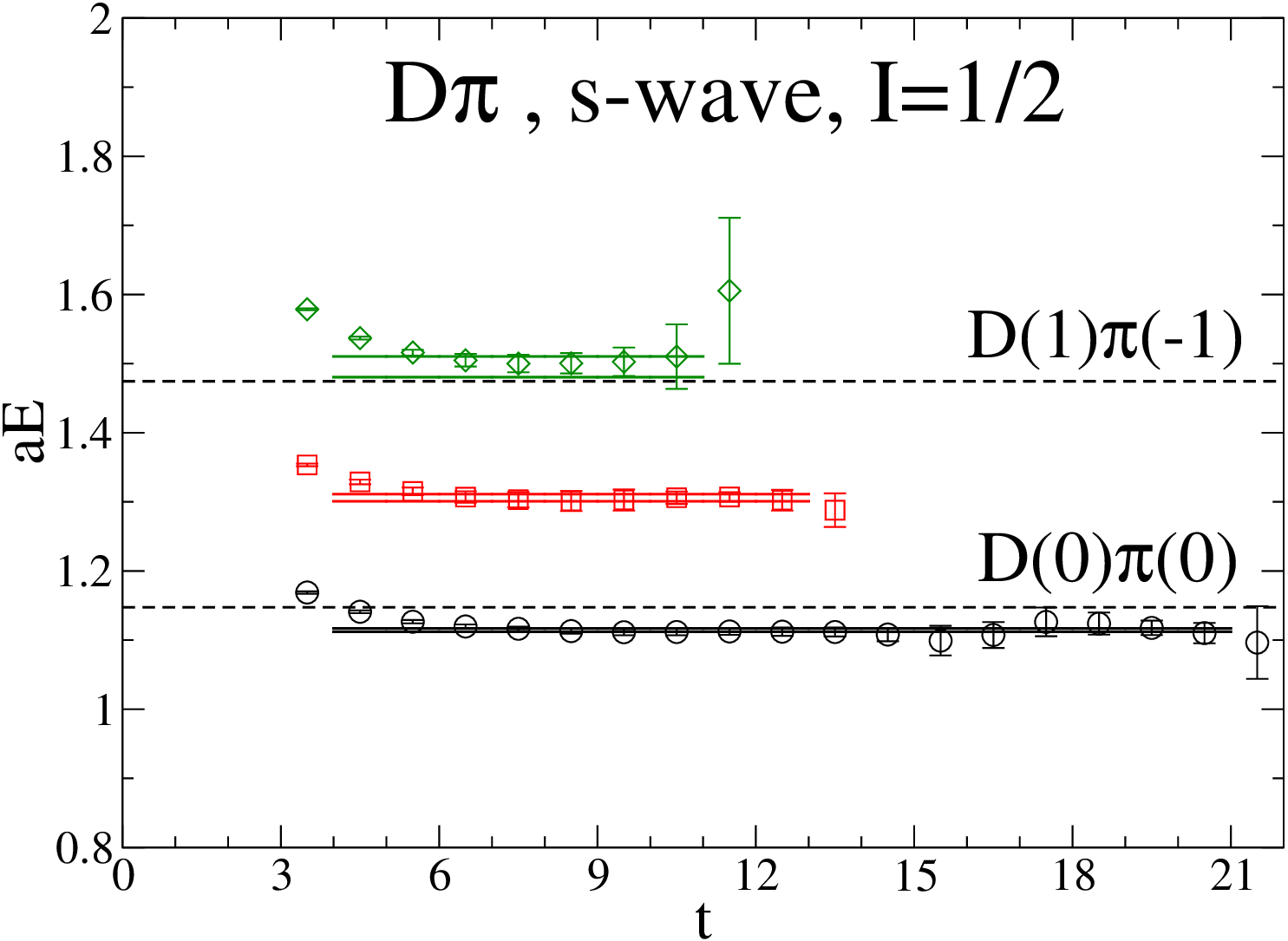}$\quad\qquad$ 
\includegraphics[width=42mm,clip]{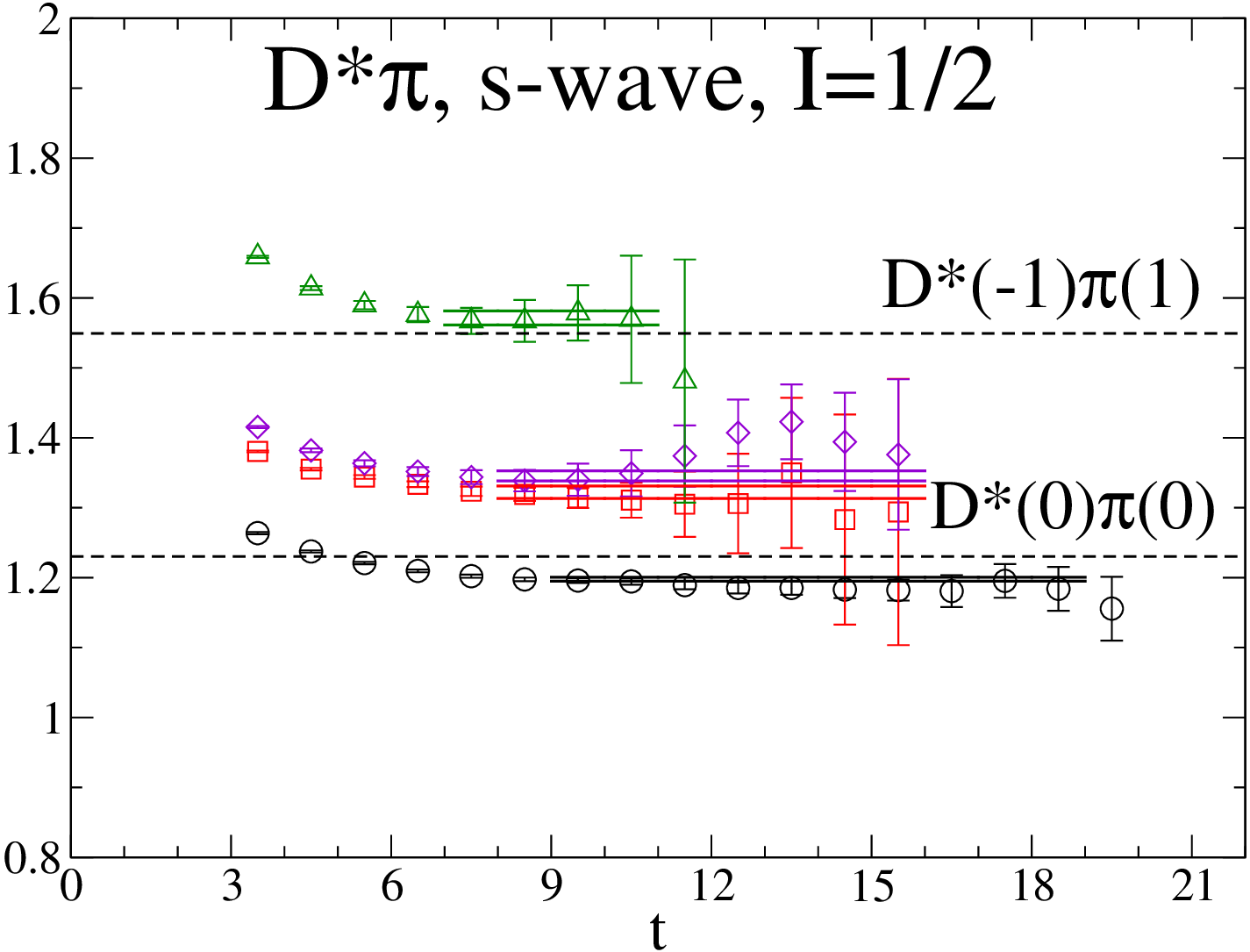}
\caption{The energy levels (effective masses of eigenvalues) for $K\pi$ ($J^P=0^+,~1^-$), $D\pi$ ($J^P=0^+$) and $D^*\pi$ ($J^P=1^+$) scattering with $\vec P=0$ \cite{Lang:2012sv,Mohler:2012na}. Dashed lines indicate energies of non-interacting scattering states. } 
\label{fig:eff}
\end{center}
\end{figure}

The scattering energy levels (black and green in Fig. \ref{fig:eff}) appear at  $E(L)=\sqrt{m_1^2+\vec p_1^{~2}}+\sqrt{m_2^2+\vec p_2^{~2}}+\Delta E(L)$ with  discrete $\vec p_i=\vec n\tfrac{2\pi}{L}$ due to the periodic boundary conditions in space. 
The energy shift $\Delta E(L)$ in the finite volume is due to the strong interaction of the two mesons. The negative shift  of the lowest level for $I=1/2$ s-wave scattering of $K\pi$, $D\pi$ and $D^*\pi$ in Fig. \ref{fig:eff}
indicates  attractive interaction. The positive shift for  $K\pi$ with $I=3/2$ indicates repulsive interaction. In addition the the scattering levels shown in black and green, the presence of resonances in $I = 1/2$ channels leads to the
   levels drawn in red and violet.  These indicate s-wave resonances $D_0(2400)$ in $D\pi$, $D_1(2420)$ and $D_1(2430)$ in $D^*\pi$, $K_0(1430)$ in $K\pi$, as well as  p-wave resonances $K^*(892)$, $K^*(1410)$, $K^*(1680)$ in $K\pi$ \cite{Mohler:2012na,Lang:2012sv}. We do not find an additional level due to $\kappa$ \cite{Lang:2012sv}, which is in agreement with the fact that the experimental phase shift does not reach $90^\circ$ below $1$ GeV.   We  also do not find  additional energy levels in $I=3/2$ channels, in line with absence of the exotic resonances in experiment. 

\vspace{-1cm}

\section{Phase shifts and resonance parameters}

The energy shift in finite volume  reveals the attractive or repulsive nature of the interaction. However, it  also rigorously renders the  phase shift for the elastic scattering in the infinite volume via the L\"uscher's relation. In particular, the energy level $E(L)$ for a scattering system with momenta $\vec P$ renders the elastic phase shift $\delta(s)$ at $s=E^2-\vec P^2$ in partial wave $l$. Note that the extraction of $\delta(s)$ is straightforward only when the partial-wave mixing due to the discrete symmetry is absent or negligible, which 
usually holds well for $\vec P=0$, but  holds rarely  for the scattering of two particles with different mass and $\vec P\not =0$ \cite{Leskovec:2012gb}. 

\begin{figure}[htb]
\begin{center}
\includegraphics[width=40mm,clip]{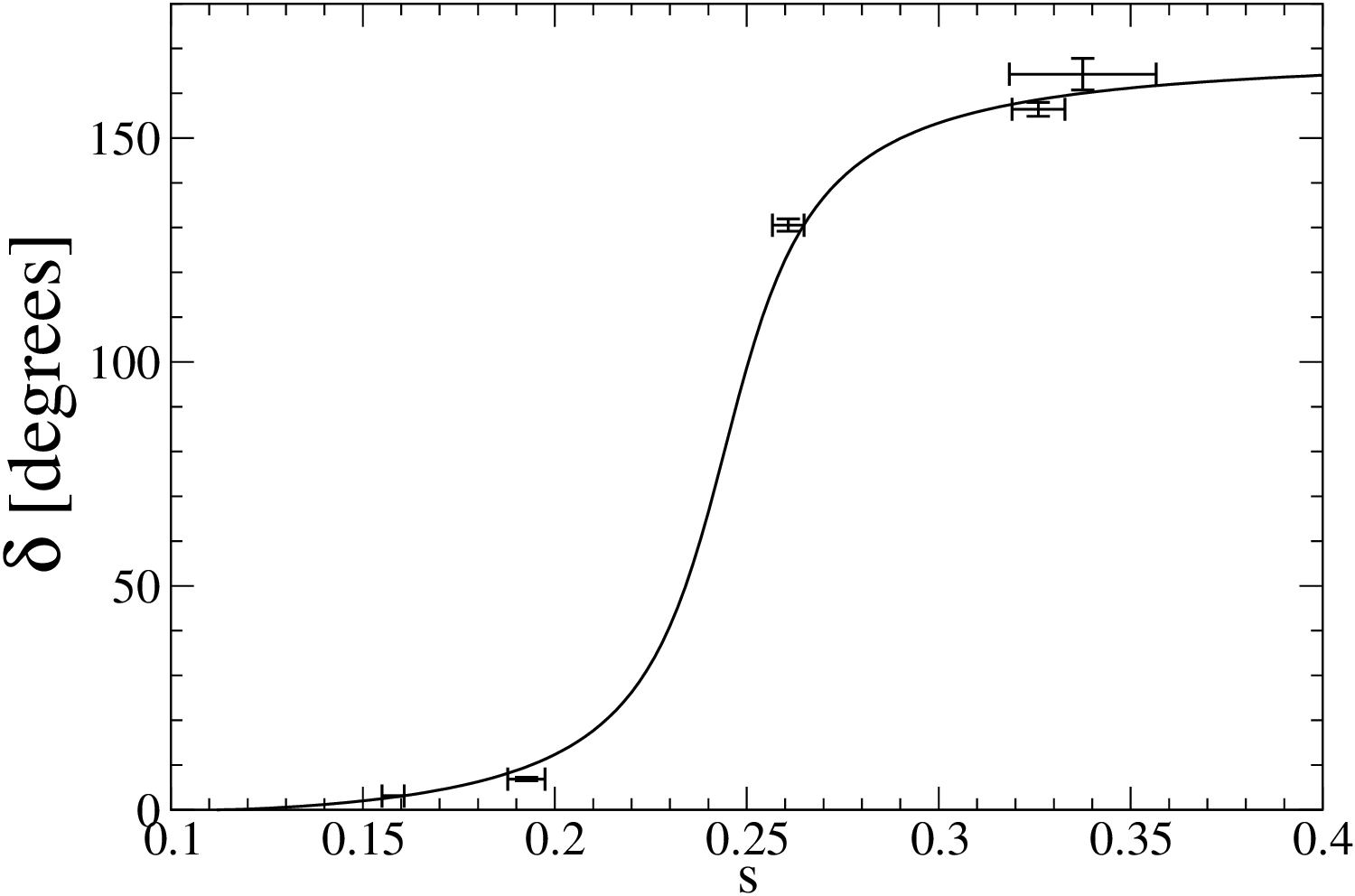}$\qquad$
\includegraphics[width=38mm,clip]{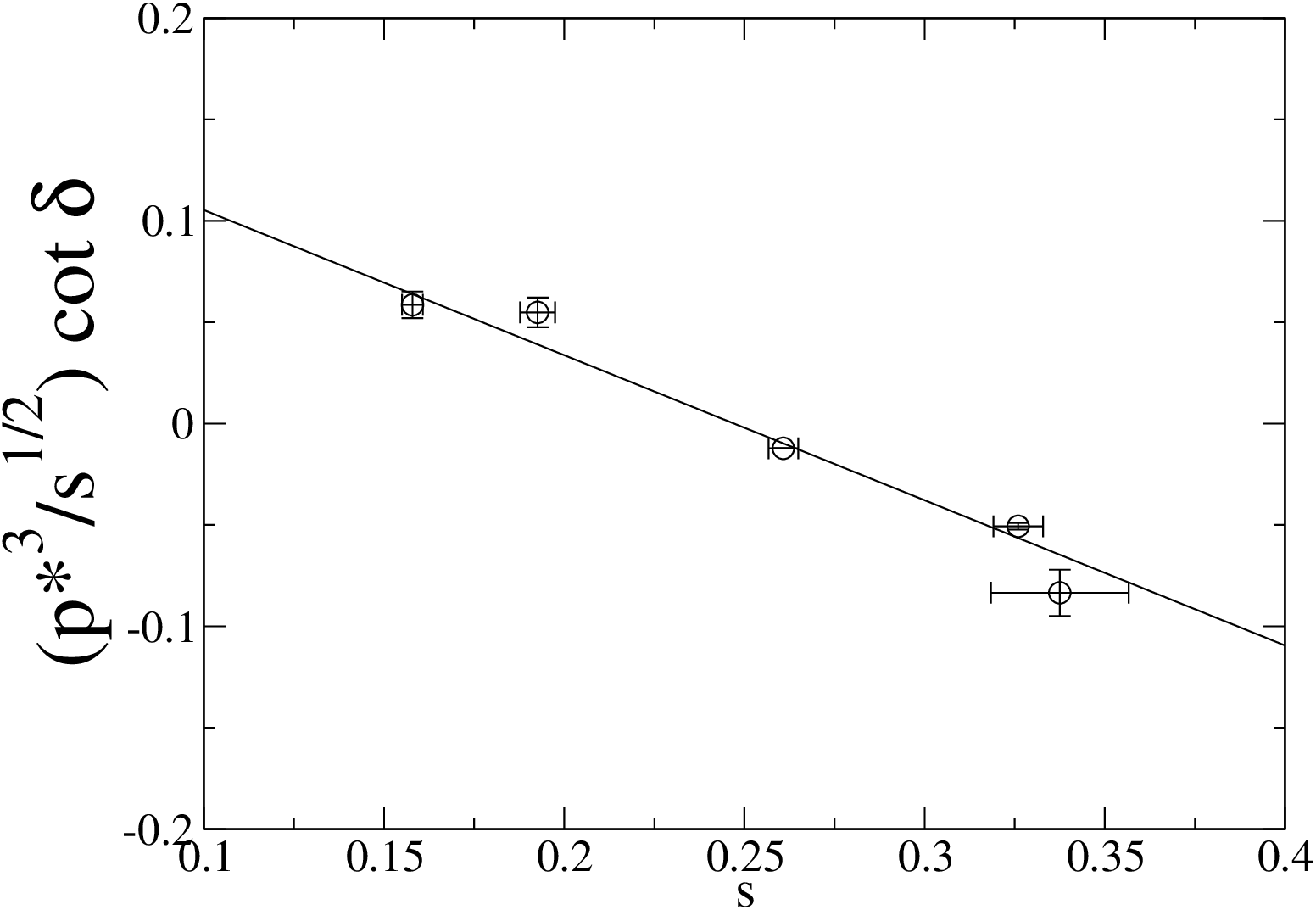}
\caption{Left: the p-wave $\pi\pi\to\pi\pi$ phase shift $\delta$  with $I=1$ \cite{Lang:2011mn}. Right: the  corresponding  $(p^*)^3\cot\delta/\sqrt{s}$ (\ref{BW_xy}). } 
\label{fig:rho}
\vspace{-1cm}
\end{center}
\end{figure}

The $\pi\pi\to\rho\to\pi\pi$ was simulated for three different  $\vec P$ in \cite{Lang:2011mn} and  five energy levels  lead to resonant phase shift  in Fig. \ref{fig:rho}.  The linear fit of the resulting $p^{*3}\cot \delta /\sqrt{s}$ (\ref{BW_xy}) leads to $g_{\rho \pi\pi}^{lat}\equiv \sqrt{6\pi}g=5.13\pm 0.20$ and $m_{\rho}^{lat}=792\pm 10~$MeV compared to $g_{\rho\pi\pi}^{exp}=5.97$ and $m_{\rho}^{exp}=775~$MeV. This is the only resonance where proper lattice treatment 
 has reached a certain level of maturity (see \cite{Lang:2011mn,Pelissier:2012pi,Dudek:2012xn} and references therein). A particularly detailed  and impressive shape of the resonant phase shift curve was achieved in \cite{Dudek:2012xn}  at a  heavier pion mass  $m_\pi\simeq 400~$MeV.

%\vspace{0.1cm} 

Over the past year we performed the first simulation of the  $K\pi$ \cite{Lang:2012sv}, $D\pi$, $D^* \pi$ \cite{Mohler:2012na} and $\rho \pi$ \cite{Prelovsek:2011im} scattering and the corresponding resonances. Since this involves scattering of two particles with different masses, we considered only  $\vec P=0$ when the mixing of different $l$ is absent or negligible \cite{Leskovec:2012gb}. 

\begin{figure*}[htb]
\begin{center}
\includegraphics*[width=0.35\textwidth,clip]{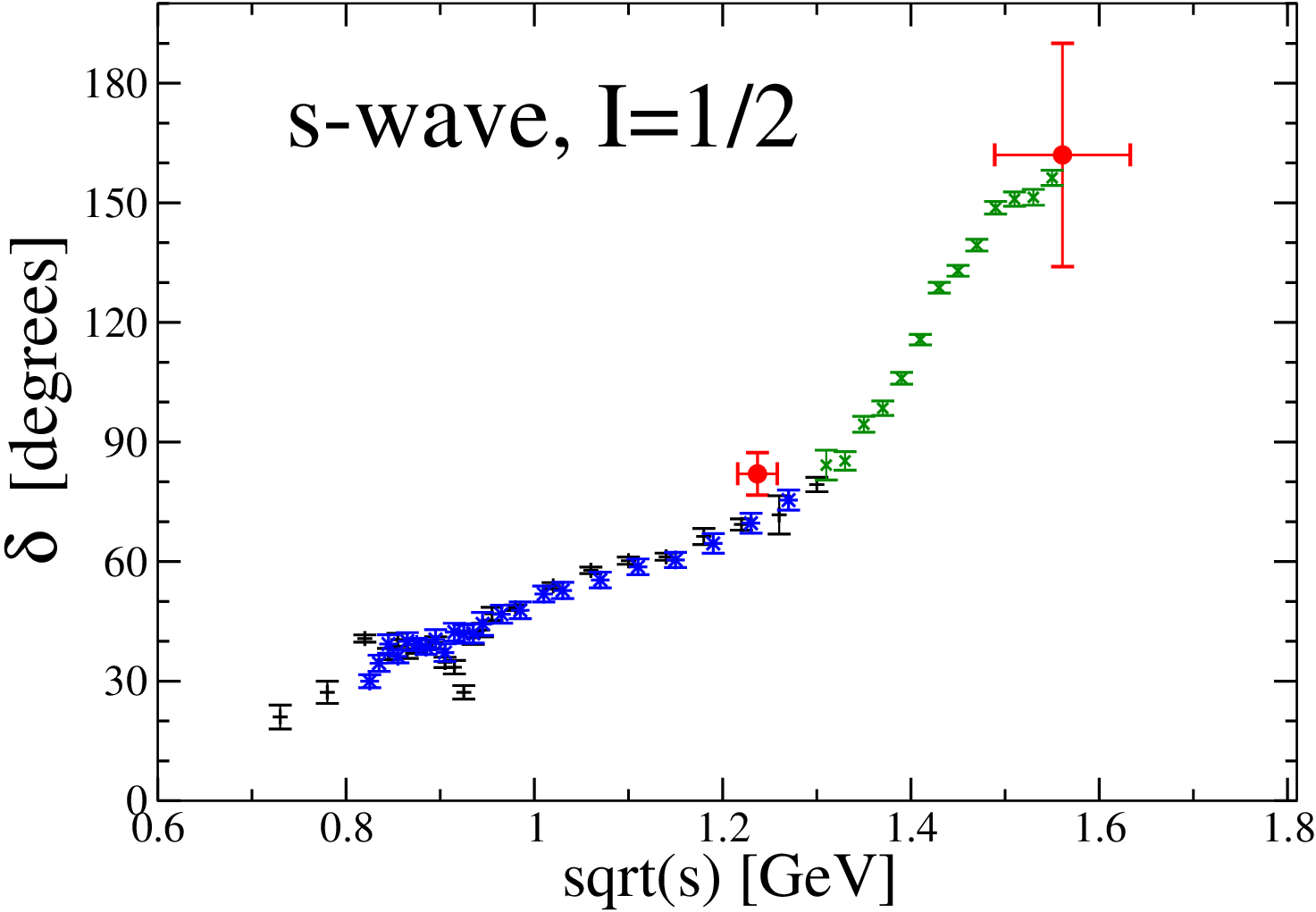}
\includegraphics*[width=0.35\textwidth,clip]{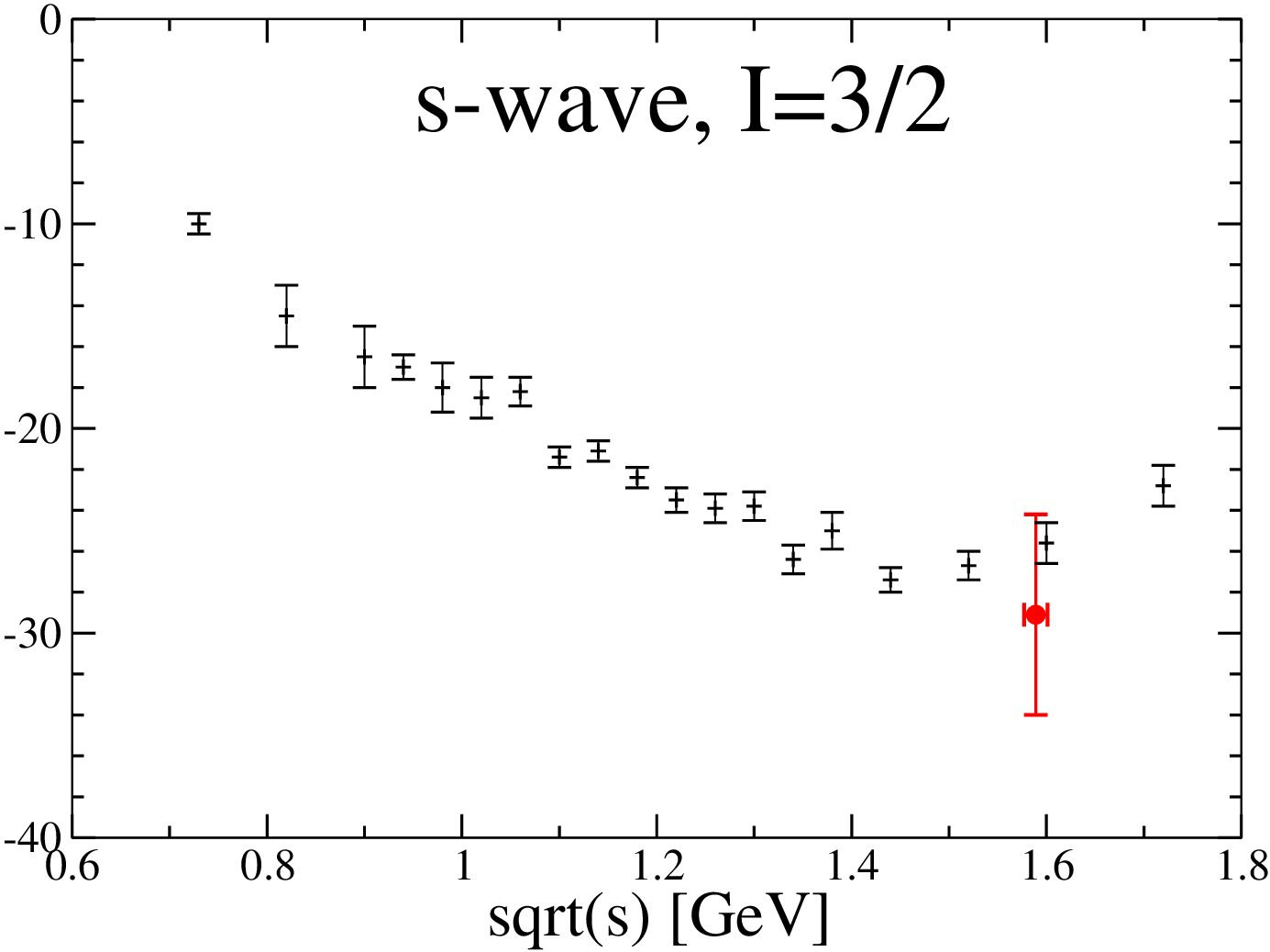}
\includegraphics*[width=0.35\textwidth,clip]{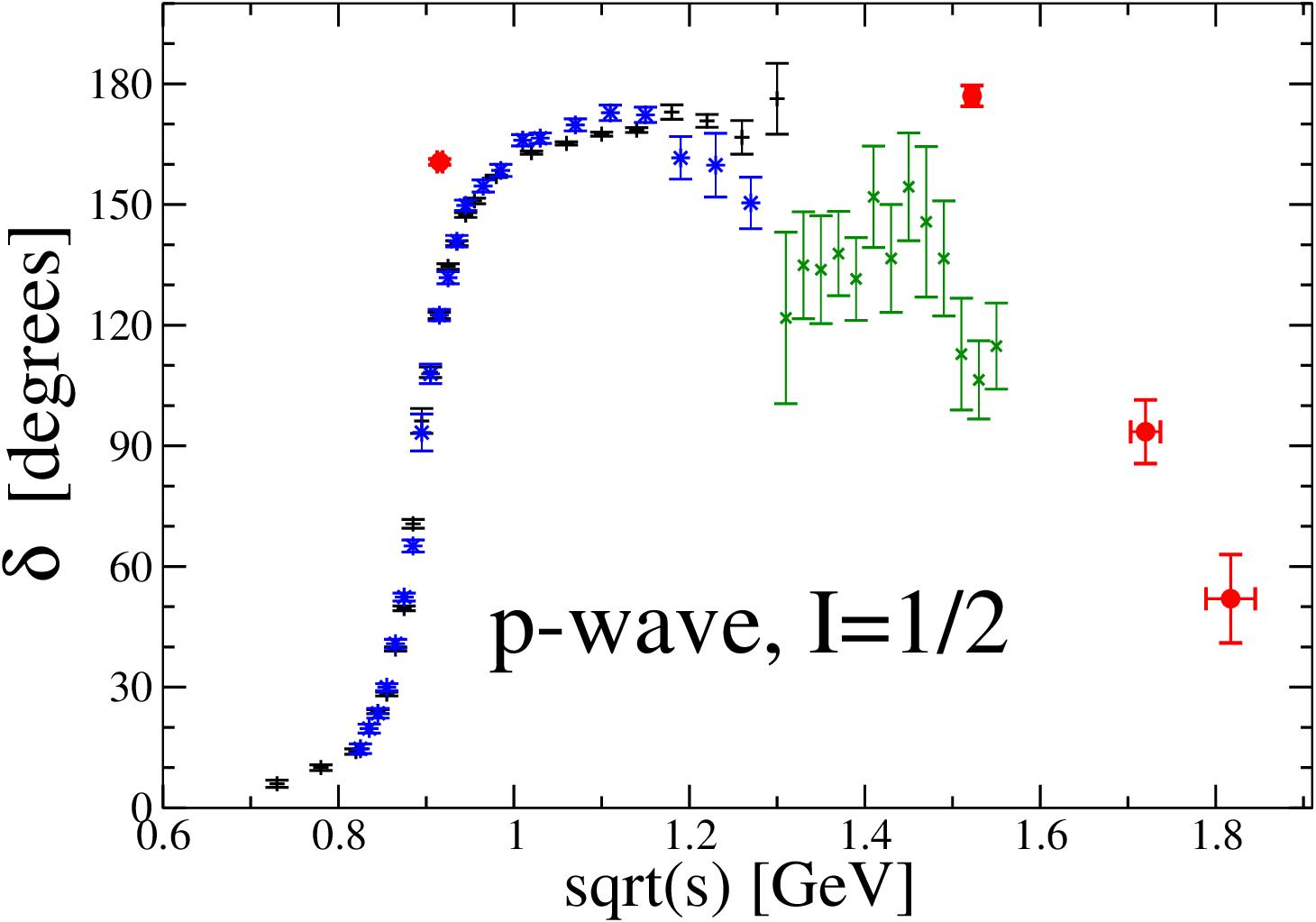}
\includegraphics*[width=0.35\textwidth,clip]{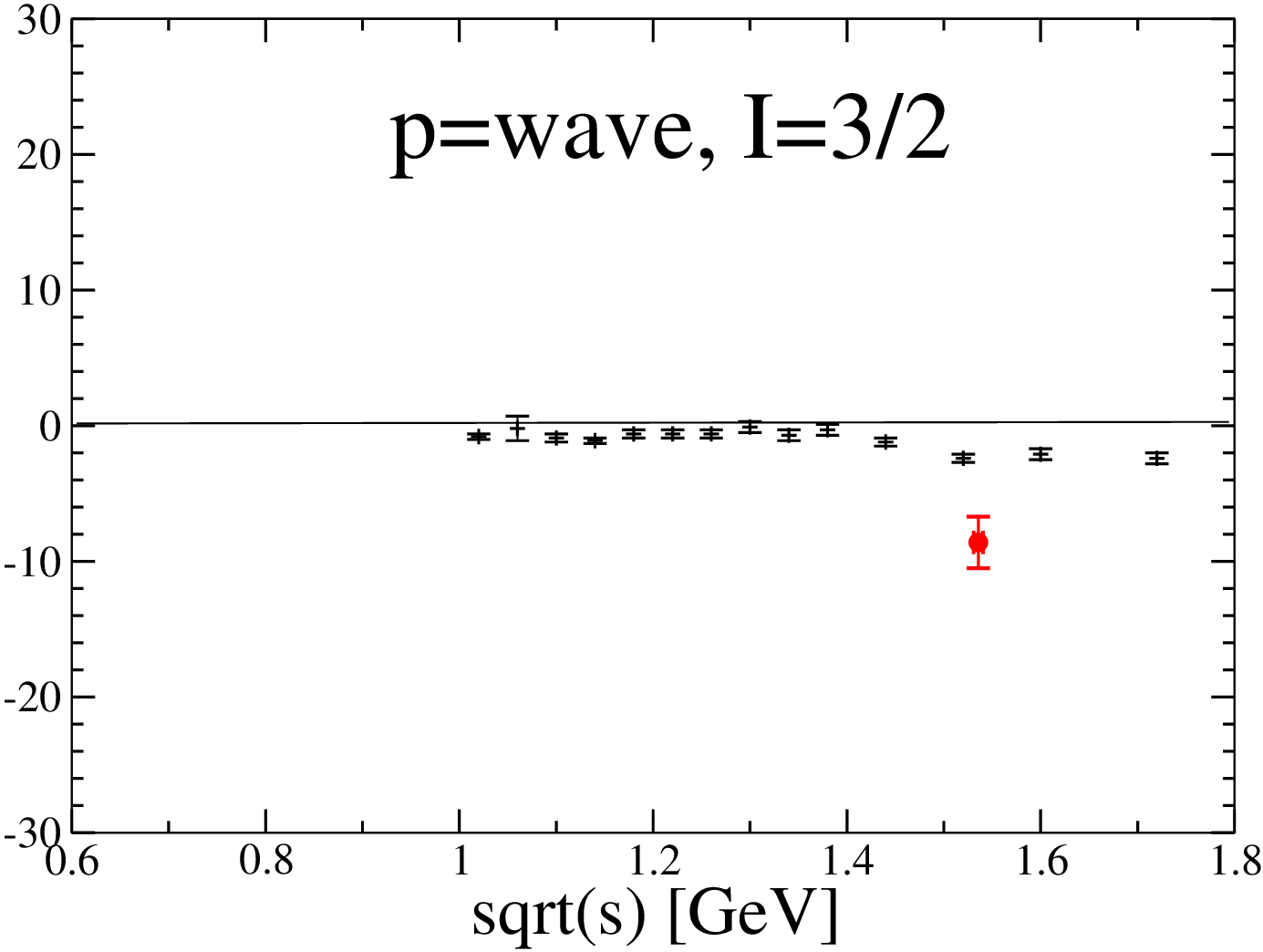}
\end{center}
\caption{
 The  $K\pi$  phase shifts $\delta_\ell^I$  in  channels $l=0,1$ and $I=1/2,~3/2$ as a function of  $K\pi$ invariant mass $\sqrt{s}$. The lattice results are given by the red circles (they apply for $m_\pi\!\simeq \!266~$MeV) \cite{Lang:2012sv}, while the other points are experimental phase shifts. }\label{fig:Kpi_phases}
\end{figure*}

   The energy levels for $K\pi$ scattering in Fig. \ref{fig:eff} lead to the phase shifts in Fig. \ref{fig:Kpi_phases} for s-wave and p-wave with $I=1/2,~3/2$. These are in qualitative agreement with the experimental ones. 

Next we concentrate on the  charmed resonances that appear in $D\pi$ and $D^*\pi$.
The  three energy levels for $D\pi$ s-wave scattering in Fig. \ref{fig:eff} lead to the phase shifts \cite{Mohler:2012na}, and the linear fit (\ref{BW_xy}) over three points leads to $m$ and $\Gamma$ (or $g$)  for the broad $D_0^*(2400)$  in Table \ref{tab:D}. The analysis of $D^*\pi$ spectrum with $J^P=1^+$ is more complicated since there are two nearby  resonances in experiment, as evidenced also by the red and violet levels in Fig. \ref{fig:eff}. 
We find that the  red level for $D^*\pi$ scattering  is due to the narrow $D_1(2420)$ which decays only in d-wave in the $m_c\to \infty$ limit. In this limit the remaining three levels are related to s-wave $D^*\pi$ scattering which is dominated by the broad $D_1(2430)$. A linear fit  (\ref{BW_xy}) through these three points leads to the mass and the width for the broad $D_1(2430)$  in Table \ref{tab:D}. 

The resulting masses and widths of $D_0^*(2400)$ and $D_1(2430)$ agree quite well with the experimental ones. Since $D_0^*(2400)$, located at $\simeq 2318~$MeV, is very close to its strange partner $D_{s0}^*(2317)$, several authors proposed that $D_0^*(2400)$  has a sizable tetraquark component $\bar c\bar ssu$. We get $m_{D_0^*(2400)}$ near the experimental value without explicitly incorporating the additional strange valence  pair\footnote{ The $\bar ss$ can  not appear as intermediate state in our simulation without dynamical $s$.}. 

{\small 
\begin{table}[htb]
\begin{tabular}{c|cc|cc}
&$m_{D_0^*(2400)}-\bar m$ & $g_{ ~D_0^*(2400) \to D\pi}$ & $m_{D_1(2430)}-\bar m$ & $g_{ ~D_1(2430) \to D\pi}$ \\
\hline
lat\cite{Mohler:2012na} & $351\pm 21~$MeV & $2.55\pm 0.21~$GeV & $381\pm 20~$MeV & $2.01 \pm 0.15~$GeV \\
exp & $347\pm 29~$MeV & $1.92\pm 0.14~$GeV & $456\pm 40~$MeV & $2.50\pm 40~$GeV
\end{tabular}
\caption{  The charmed resonance masses (with respect to $\bar m\equiv \tfrac{m_D+3m_{D^*}}{4}$) and the couplings $g$, which parametrize the  widths $\Gamma=g^2p^*/s$. The experimental couplings $g$ are derived from total widths. }
\label{tab:D}
\end{table}
}

\begin{figure}[tbh]
\begin{center}
\includegraphics[width=65mm,clip]{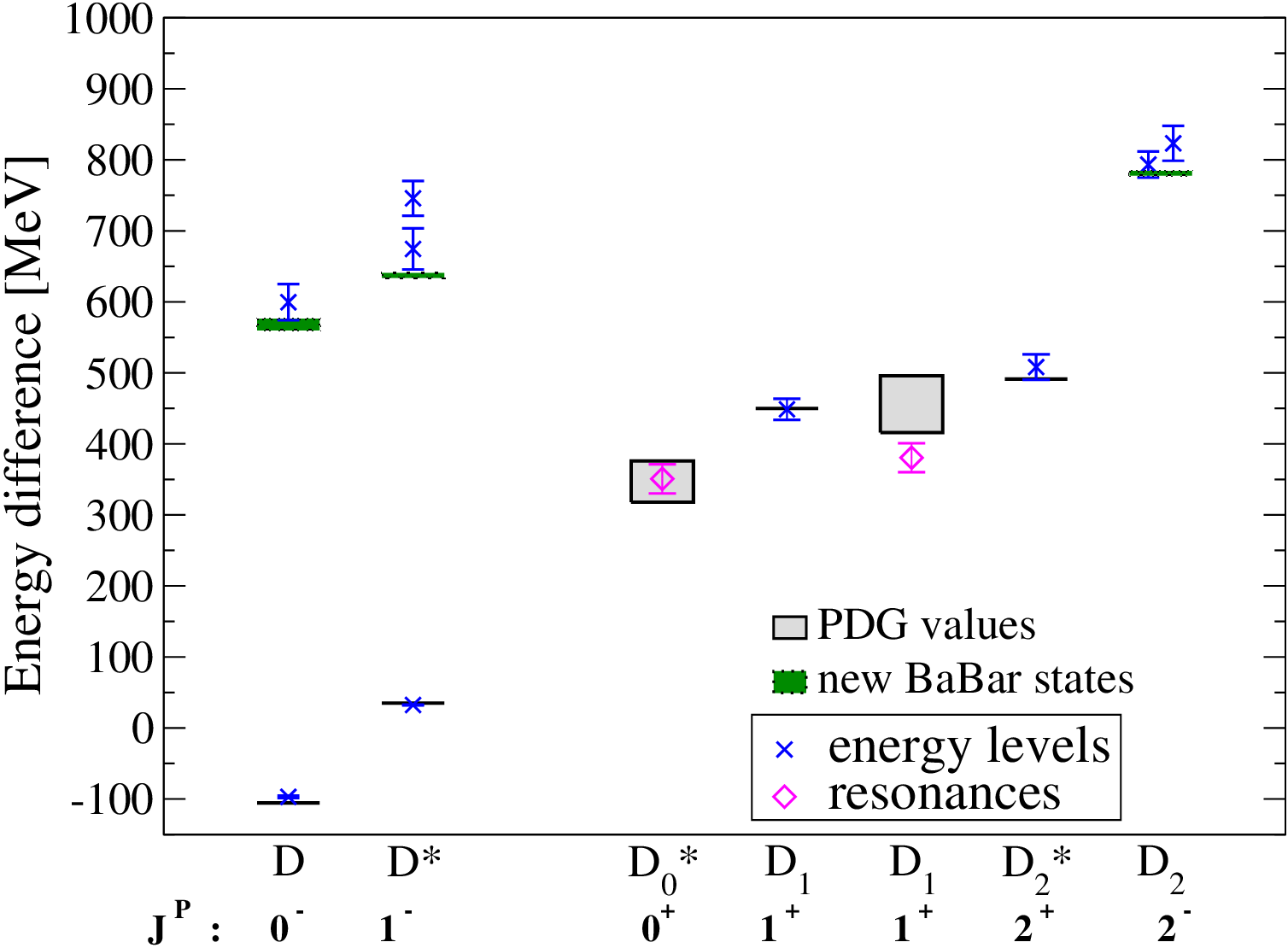}
\caption{Energy differences  $m-\tfrac{1}{4}(m_{D}+3m_{D^*})$ for $D$ mesons on lattice  \cite{Mohler:2012na} and in  experiment; the reference  mass is $\bar m=\tfrac{1}{4}(m_{D}+3m_{D^*})\approx 1971~$MeV in experiment. Magenta diamonds give masses for states simulated as resonances \cite{Mohler:2012na}. Masses extracted as energy levels on a finite lattice are displayed as blue crosses \cite{Mohler:2012na}.} 
\label{fig:D}
\end{center}
\end{figure}

{\normalsize
The compilation of the  $D$ meson spectrum  in Fig. \ref{fig:D}  shows quite good agreement with  experiment \cite{Mohler:2012na}.  The masses of broad resonances $D_0^*(2400)$ and $D_1(2430)$ are extracted as explained above. Other four low-lying $J^P=0^-,~1^-,1^+,~2^+$ states are stable or very narrow, so they were simulated using  ${\cal O}=\bar c\Gamma u$  and  $m\!=\!E$ ($\vec P\!=\!0$) is employed like in all previous simulations. This narrow-width approximation is applied also for the excited states in $J^P=0^-,~1^-,~2^-$ channels, which are compared to the states observed by BaBar in 2010 \cite{delAmoSanchez:2010vq}; unfortunately these are not yet confirmed by any other experiment. 

The scattering lengths for $K\pi$, $D\pi$ and $D^*\pi$ were also extracted in  \cite{Lang:2012sv,Mohler:2012na}. 
 
\vspace{0.1cm}

In conclusion, $\rho$ is the only resonance that was treated properly  by several lattice groups up to now. We presented the first results of the strange and charmed resonances based on the simulation of the corresponding scattering channels. 

\vspace{0.2cm}
%{\bf Acknowledgments}

We thank Anna Hasenfratz for providing the gauge configurations. Fermilab is operated by Fermi Research Alliance, LLC under Contract No.
De-AC02-07CH11359 with the United States Department of Energy.
}
\bibliographystyle{h-physrev4}
\bibliography{bibtex_excitedQCD13}

\begin{thebibliography}{10}

\bibitem{Lang:2011mn}
C.~B. Lang, D.~Mohler, S.~Prelovsek and M.~Vidmar,
\newblock Phys.Rev. {\bf D84}, 054503 (2011).

\bibitem{Pelissier:2012pi}
C.~Pelissier and A.~Alexandru,
\newblock Phys.Rev. {\bf D87}, 014503 (2013).

\bibitem{Dudek:2012xn}
J.~J. Dudek, R.~G. Edwards and C.~E. Thomas,
\newblock Phys.Rev. {\bf D87}, 034505 (2013).

\bibitem{Mohler:2012na}
D.~Mohler, S.~Prelovsek and R.~M. Woloshyn,
\newblock Phys. Rev. D {\bf 87}, 034501 (2013).

\bibitem{Lang:2012sv}
C.~B. Lang, L.~Leskovec, D.~Mohler and S.~Prelovsek,
\newblock Phys.Rev. {\bf D86}, 054508 (2012).

\bibitem{Mohler:2012nh}
D.~Mohler,
\newblock PoS {\bf LATTICE2012}, 003 (2012), [arXiv:1211.6163].
%%CITATION = ARXIV:1211.6163;%%

\bibitem{Peardon:2009gh}
M.~Peardon {\em et~al.},
\newblock Phys. Rev. D {\bf 80}, 054506 (2009).

\bibitem{Leskovec:2012gb}
L.~Leskovec and S.~Prelovsek,
\newblock Phys.Rev. {\bf D85}, 114507 (2012).

\bibitem{Prelovsek:2011im}
S.~Prelovsek, C.~B. Lang, D.~Mohler and M.~Vidmar,
\newblock PoS {\bf LATTICE2011}, 137 (2011), [arXiv:1111.0409].
%%CITATION = ARXIV:1111.0409;%%

\bibitem{delAmoSanchez:2010vq}
The BABAR, P.~del Amo~Sanchez {\em et~al.},
\newblock Phys. Rev. {\bf D82}, 111101 (2010).

\end{thebibliography}

\end{document}